\documentclass[11pt,a4paper]{article}
\usepackage[margin=24mm]{geometry}
\usepackage[T1]{fontenc}
\usepackage[utf8]{inputenc}
\usepackage{lmodern,amsmath,amssymb,mathtools,bm,graphicx,booktabs,hyperref,xurl}
\graphicspath{{source_archive/}}
\hypersetup{colorlinks=true,linkcolor=blue,citecolor=blue,urlcolor=blue}

\title{Quantum Sensitivity of $N$-Component Cat States}

\author{
	Ranjit Singh$^{1}$ and Alexander E. Teretenkov$^{2}$\\[2mm]
	\small $^1$Independent Researcher, Domodedovo, Moscow Region, Russia\\
	\small \texttt{ranjit.singh@mail.ru}\\[2mm]
	\small $^2$Department of Mathematical Methods for Quantum Technologies,\\
	\small Steklov Mathematical Institute of the Russian Academy of Sciences,\\
	\small 8 Gubkina St., Moscow, 119991, Russia\\
	\small \texttt{taemsu@mail.ru}
}

\date{}
\begin{document}
\maketitle
\begin{abstract}
We extend our previous study of squeezed two-component Schr\"odinger cat states to a phase-referenced single-mode family with an arbitrary number $N$ of coherent components and general relative phases. Within the established circular-state construction, we derive a closed complex overlap kernel for the ordered action of displacement, phase rotation, and squeezing, retaining exact finite-amplitude normalization. We distinguish finite-displacement orthogonality from local phase-estimation sensitivity quantified by quantum Fisher information (QFI). For equal-phase states, we compare $N=2,\ldots,9$ at a common mean photon number and include coherent-state and squeezed-vacuum benchmarks. A bound on the higher photon-number sectors establishes the fixed-energy asymptote $F_{Q,N}=4\bar n(N-\bar n)+o(1)$. Its leading term agrees with the known QFI of a vacuum--$N$-photon superposition. The resulting decrease of the ideal local quantum Cram\'er--Rao bound does not establish unlimited attainable precision: prior phase information, sample size, phase-reference resources, optical loss, state preparation, and measurement choice remain relevant.
\end{abstract}
\section{Introduction}
Non-Gaussian superpositions of coherent states display interference in optical phase space and offer resources for sensing small transformations. The even and odd coherent states~\cite{Dodonov1974} are two-component examples of the broader class of circular coherent-state superpositions~\cite{Janszky1993}. Their photon-number and phase-space properties have been studied for general component numbers~\cite{MizrahiJose2002,LanXu2022}. Here we extend our own analysis of squeezed two-component cat-state sensitivity~\cite{ST} to the finite-amplitude, phase-referenced single-mode problem with arbitrary $N$ and relative phases. A review of optical quantum metrology and its experimental limitations is given in Ref.~\cite{Alodjants2024}.

Displacement-induced distinguishability of cat and compass states is connected with sub-Planck interference structures~\cite{Zurek2001,Toscano2006}. More recent work has examined directional sensitivity and its approach to isotropy for superpositions of rotated compass states~\cite{Shukla2023}, as well as displacement overlaps of circular cats with larger even component numbers and their optomechanical analogues~\cite{Hailin2025}. Following the displacement-based analysis in our earlier study~\cite{ST}, we use $|\langle\psi|D(\delta)|\psi\rangle|^2$ for a normalized pure state to characterize finite transformations. Its zeros determine orthogonality, whereas the quantum Fisher information (QFI) independently quantifies local phase-estimation sensitivity; these measures are complementary but not equivalent.

Multicomponent cats have also been investigated as resources for phase estimation. Lee and coauthors considered four-headed cats and extensions to more components in two-mode interferometry~\cite{Lee2015}, and subsequently optimized phase estimation with generalized circular cats having different photon-number residue sectors~\cite{Lee2020}. Bibi and coauthors compared two-, three-, and four-headed cats for phase and quadrature sensing~\cite{Bibi2026}; photon-engineered multiheaded probes have also been studied~\cite{BibiIqbal2026}. These settings motivate an explicit distinction between the component amplitude, the mean energy of the normalized probe, and the phase generator. In this work the parameter is encoded by the single-mode operator $e^{-i\theta n}$, and the equal-energy comparison adjusts the amplitude separately for each $N$.

A mean-photon-number constraint alone does not bound phase QFI. Earlier constructions demonstrate this for fluctuating-photon-number interferometry~\cite{Zhang2013} and for single-mode probes with large photon-number variance~\cite{Lee2019}. In particular, a vacuum--$N$-photon superposition has QFI $4\bar n(N-\bar n)$ at fixed mean photon number~\cite{Lee2019}. Our large-$N$ result establishes how the circular-cat family approaches this value, with an explicit bound on the contribution of higher sectors. The distinction between local QFI and operational phase uncertainty is essential in interpreting such limits~\cite{RivasLuis2012,Berry2012,Pezze2013}.

Possible generation mechanisms for cat-like non-Gaussian states have been investigated in our earlier studies involving $\chi^{(2)}$ spontaneous parametric down-conversion~\cite{STPLA} and $\chi^{(3)}$ microresonators~\cite{SinghTeretenkov2026Microring,SinghTeretenkov2026Kerr}. These studies motivate nonlinear platforms, but do not establish preparation of the arbitrary ideal pure states assumed here. Independently, three- and four-component superpositions have been generated in circuit QED~\cite{Vlastakis2013}, and generalized-parity measurement protocols address the preparation of larger multicomponent cats while tracking imperfections and success probabilities~\cite{Zeytinoglu2026}.

We first recall the photon-number and Wigner representations needed for the calculation, distinguishing established circular-state properties from the results developed here. We then evaluate the complex overlap for the ordered transformation $S^\dagger U(\theta)D(\delta)S$, including general relative phases and finite-amplitude normalization. For equal phases, we examine $N=2,3,4$ explicitly, compare phase QFI for $N=2,\ldots,9$ at fixed mean energy, and prove the large-$N$ asymptote by controlling the photon-number tail. We also compare with the optimal pure single-mode Gaussian probe~\cite{Monras2006} and state the reference, measurement, and statistical assumptions needed for a local-estimation interpretation.

\section{Symmetric multicomponent coherent states}
For $N\ge2$, let $\alpha_k=\alpha e^{2\pi i k/N}$. We use the established circular coherent-state construction~\cite{Janszky1993,MizrahiJose2002,LanXu2022}, first allowing arbitrary relative phases $\phi_k$:
\begin{equation}\label{state}
|\Psi_N\rangle=\mathcal N_N\sum_{k=0}^{N-1}e^{i\phi_k}|\alpha_k\rangle,
\qquad
\mathcal N_N^{-2}=\sum_{j,k=0}^{N-1}e^{i(\phi_k-\phi_j)}
\exp[-|\alpha|^2+\alpha_j^*\alpha_k].
\end{equation}
The normalization is positive whenever the unnormalized superposition is nonzero. For the plotted examples and simplified expressions below we subsequently set $\phi_k=0$. For this equal-phase family, writing $r=|\alpha|^2$ and $\zeta_N=e^{2\pi i/N}$ gives
\begin{equation}\label{fock}
|\Psi_N\rangle=\frac1{\sqrt{Z_N(r)}}\sum_{m=0}^\infty\frac{\alpha^{Nm}}{\sqrt{(Nm)!}}|Nm\rangle,
\quad Z_N(r)=\frac1N\sum_{k=0}^{N-1}e^{r\zeta_N^k},\quad
\mathcal N_N=\frac{e^{r/2}}{N\sqrt{Z_N(r)}}.
\end{equation}
Only multiples of $N$ occur in the Fock expansion, as in earlier circular-state representations~\cite{MizrahiJose2002,LanXu2022}. This selection rule changes the photon-number distribution as $N$ changes, even when $|\alpha|$ is held fixed. We reproduce the derivation in Appendix~\ref{app:norm} to fix the normalization and the moments used below.
\subsection{Explicit equal-phase states for \texorpdfstring{$N=2,3,4$}{N=2,3,4}}
The first three families have particularly transparent coherent-state representations:
\begin{align}
|\Psi_2\rangle&=\mathcal N_2(|\alpha\rangle+|-\alpha\rangle),\\
|\Psi_3\rangle&=\mathcal N_3(|\alpha\rangle+|\alpha e^{2\pi i/3}\rangle+|\alpha e^{4\pi i/3}\rangle),\\
|\Psi_4\rangle&=\mathcal N_4(|\alpha\rangle+|i\alpha\rangle+|-\alpha\rangle+|-i\alpha\rangle).
\end{align}
Their exact Fock expansions are given by Eq.~\eqref{fock} with $N=2,3,4$; hence only $|2m\rangle$, $|3m\rangle$, or $|4m\rangle$, respectively, occur. The even cat is $|\Psi_2\rangle$; the odd cat instead has a relative phase $\pi$ and belongs to the general-phase family in Eq.~\eqref{state}. More generally, choosing $\phi_k=-2\pi qk/N$, with integer $q\in\{0,\ldots,N-1\}$, selects the Fock sectors $n\equiv q\pmod N$, as in the generalized cats of Ref.~\cite{Lee2020}. In particular, odd numbers of coherent components ($N=3,5,\ldots$) are included in Eq.~\eqref{state}; an \emph{odd number of components} must not be confused with \emph{odd photon-number parity}. The relative phases specify the superposition and its photon-number sectors.
\subsection{Wigner representation}
\begin{figure}[htbp]\centering
\includegraphics[width=\textwidth]{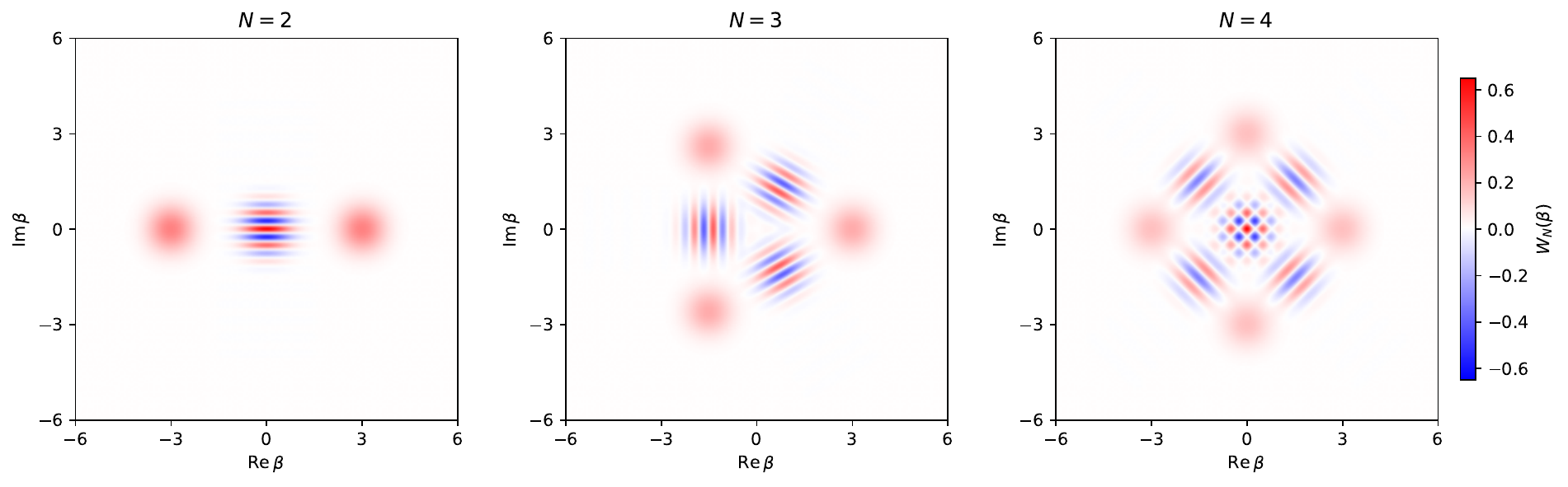}
\caption{Wigner functions for equal-phase $N=2,3,4$ cats at $\alpha=3$, calculated from Eq.~\eqref{wignerexact}. The horizontal and vertical axes are $\operatorname{Re}\beta$ and $\operatorname{Im}\beta$.}\label{fig:wigner}
\end{figure}

We reserve $\alpha$ for the state amplitude and use $\beta\in\mathbb C$ as the phase-space variable. With $n=a^\dagger a$, in the dimensionless optical convention,
\begin{equation}\label{wignerdef}
W_N(\beta)=\frac{2}{\pi}\operatorname{Tr}\!\left[|\Psi_N\rangle\langle\Psi_N|D(2\beta)(-1)^n\right].
\end{equation}
For the equal-phase family, the standard coherent-state cross terms give the finite-sum Wigner representation~\cite{LanXu2022,Hailin2025}
\begin{equation}\label{wignerexact}
W_N(\beta)=\frac{2\mathcal N_N^2}{\pi}
\sum_{j,k=0}^{N-1}\exp\!\left[-2|\beta|^2-r+2\beta\alpha_j^*+2\beta^*\alpha_k-\alpha_j^*\alpha_k\right].
\end{equation}
The sum is real, even though its individual terms need not be. Fig.~\ref{fig:wigner} illustrates the interference and rotational symmetry of the first three families. Eq.~\eqref{wignerexact} applies to every $N\ge2$.

\section{Quantum sensitivity and state orthogonality}
Displacement-based quantum sensitivity can be characterized by the distinguishability of a state and its displaced counterpart~\cite{Zurek2001,Toscano2006}. We retain the overlap criterion used in our earlier study~\cite{ST} and in subsequent analyses of multicomponent and generalized compass states~\cite{Shukla2023,Hailin2025}.
For a normalized pure state $|\psi\rangle$, define
\begin{equation}\label{orthogonalityAmplitude}
\gamma_\psi(\delta)=|\langle\psi|D(\delta)|\psi\rangle|^2,
\qquad D(\delta)=e^{\delta a^\dagger-\delta^*a}.
\end{equation}
The states are orthogonal if $\gamma_\psi=0$ and nearly orthogonal when
$\gamma_\psi$ is sufficiently small. A smaller displacement magnitude
required to reach the same orthogonality criterion indicates greater
displacement-based sensitivity. The direction of displacement and, for
near-orthogonality, the overlap threshold must be specified; see
Appendix~\ref{app:cases}. The ordinary cat is $|\Psi_N\rangle$ and the
squeezed cat is $S(t)|\Psi_N\rangle$, extending the two-component SC--SSC
comparison in Ref.~\cite{ST}.

Let $U(\theta)=e^{-i\theta n}$. Squeezing is described by $S(t)=\exp[-igt(A_2^*a^2+A_2a^{\dagger2})]$, where $A_2=|A_2|e^{i\varphi}$. We fix the operator ordering as
\begin{equation}\label{gammas}
\gamma_N(\theta,\delta,t)=
|\langle\Psi_N|S^\dagger(t)U(\theta)D(\delta)S(t)|\Psi_N\rangle|^2.
\end{equation}
For arbitrary component phases the exact result is
\begin{equation}\label{generalphasegamma}
\gamma_N(\theta,\delta,t)=\mathcal N_N^4
\left|\sum_{j,k=0}^{N-1}e^{i(\phi_k-\phi_j)}
\chi(\alpha_j,\alpha_k;\theta,\delta,t)\right|^2,
\end{equation}
where $\chi(x,y;\theta,\delta,t)=\langle x|S^\dagger UDS|y\rangle$
is derived step by step in Appendix~\ref{app:gaussian}.
At $\theta=t=0$, Eq.~\eqref{generalphasegamma} reduces to the
displacement-based overlap in Eq.~\eqref{orthogonalityAmplitude};
at $\theta=0$ it describes displaced squeezed cats. The corresponding
limits and orthogonality conditions are given in Appendix~\ref{app:cases}.
For a small unsqueezed phase rotation,
\begin{equation}\label{localSensitivity}
\gamma_N(\theta,0,0)=1-\frac{F_{Q,N}}4\theta^2+O(\theta^4),
\qquad F_{Q,N}=4\operatorname{Var}_{\Psi_N}(n).
\end{equation}
Thus finite-displacement orthogonality and local phase-estimation QFI
describe related but distinct sensing properties.

\begin{figure}[htbp]
\centering\includegraphics[width=\textwidth]{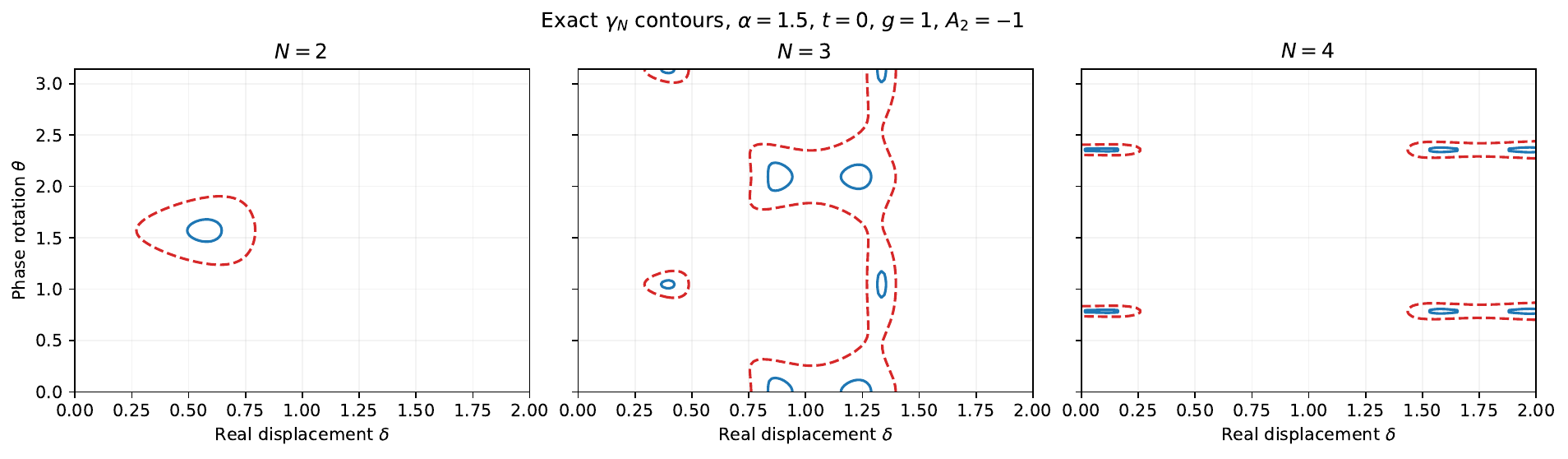}
\caption{Contours of the exact squared overlap
$\gamma_N(\theta,\delta,0)=10^{-3}$ (solid) and $10^{-2}$ (dashed)
for $N=2,3,4$, $\alpha=1.5$, and real $\delta$.
The axes are displacement $\delta$ and rotation angle $\theta$;
no squeezing is applied.}\label{fig:cont0}
\end{figure}

\begin{figure}[htbp]
\centering\includegraphics[width=\textwidth]{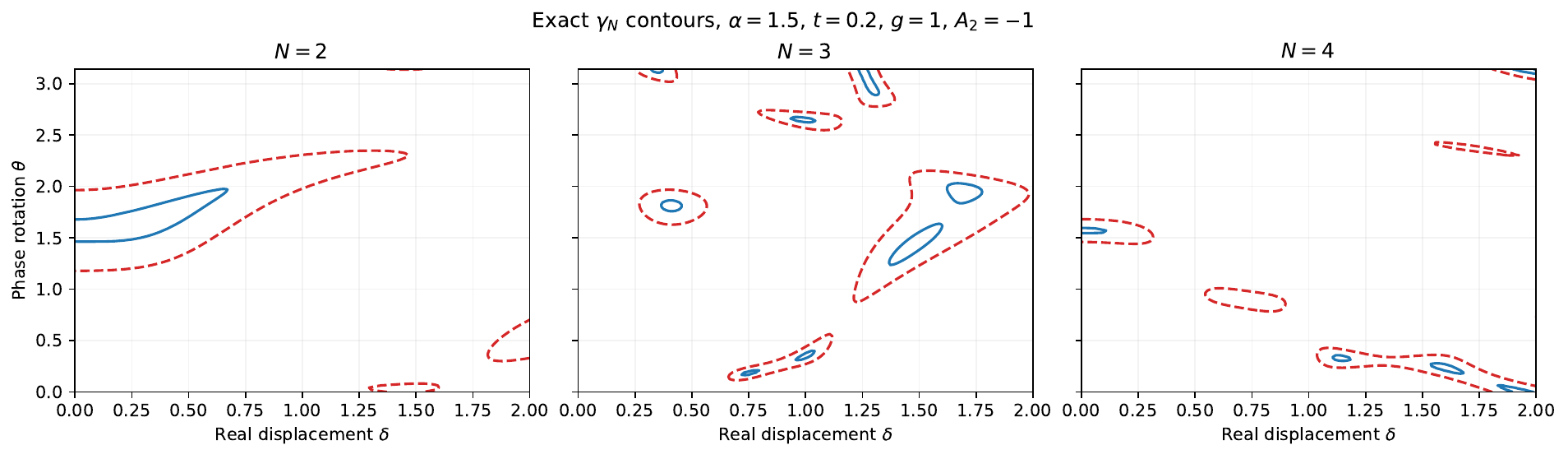}
\caption{Squared-overlap contours for squeezed states at $t=0.2$,
$g=|A_2|=1$, and $\varphi=\pi$. All other parameters and contour
levels are as in Fig.~\ref{fig:cont0}. The ordered transformation is
$S^\dagger U(\theta)D(\delta)S$.}\label{fig:cont02}
\end{figure}

For the figures we take $\phi_k=0$. The squeezed contours use $\varphi=\pi$, $g=|A_2|=1$, and $t=0.2$; the unsqueezed contours ($t=0$) are independent of $\varphi$. Figs.~\ref{fig:cont0} and \ref{fig:cont02} show near-orthogonality under combined displacement and rotation. These contours do not determine the minimum pure-displacement orthogonality scale or an estimation-uncertainty bound.

\section{Quantum Fisher Information at Fixed Mean Photon Number}
For pure-state phase encoding, the relation between statistical distance and QFI~\cite{BC,Paris} identifies the QFI with the curvature of the squared overlap at zero phase shift:
\begin{equation}\label{qfiOverlap}
F_{Q,N}
=-2\left.\frac{\partial^2\gamma_N(\theta,0,0)}
{\partial\theta^2}\right|_{\theta=0}
=4\operatorname{Var}_{\Psi_N}(\hat n).
\end{equation}
For equal-phase states, Eqs.~\eqref{fock} and \eqref{moments} yield
\begin{equation}\label{fqexact}
F_{Q,N}(r)=4\left[
r^2\frac{Z_N''(r)}{Z_N(r)}
+r\frac{Z_N'(r)}{Z_N(r)}
-r^2\left(\frac{Z_N'(r)}{Z_N(r)}\right)^2
\right].
\end{equation}
The photon-number moments and the corresponding arbitrary-phase
expression are derived in Appendix~\ref{app:norm}.
To compare phase sensitivity at equal mean photon number, we impose
\begin{equation}\label{match}
\bar n_N(r_N)=r_N\frac{Z_N'(r_N)}{Z_N(r_N)}=\bar n_\mathrm{target},\qquad \bar n_\mathrm{target}=2.25.
\end{equation}
For each $N$, Eq.~\eqref{match} has a unique positive root $r_N=|\alpha_N|^2$, as proved in Appendix~\ref{app:norm}. Table~\ref{tab:fixedmean} reports evaluations of the exact QFI formula at those roots.
\begin{table}[ht]\centering
\caption{Fixed mean photon number $\bar n=2.25$; coherent amplitudes are adjusted separately for each cat. The squeezed-vacuum benchmark is given by Eq.~\eqref{gaussianbenchmark}.}\label{tab:fixedmean}
\begin{tabular}{@{}lrrr@{}}\toprule State & $|\alpha_N|$ & $\bar n$ & $F_Q$ \\ \midrule
$N=2$ & 1.52 & 2.25 & 9.84 \\
$N=3$ & 1.51 & 2.25 & 11.50 \\
$N=4$ & 1.52 & 2.25 & 16.97 \\
$N=5$ & 1.58 & 2.25 & 25.04 \\
$N=6$ & 1.66 & 2.25 & 33.82 \\
$N=7$ & 1.74 & 2.25 & 42.77 \\
$N=8$ & 1.83 & 2.25 & 51.75 \\
$N=9$ & 1.92 & 2.25 & 60.75 \\
Coherent state & 1.50 & 2.25 & 9.00 \\
Squeezed vacuum & --- & 2.25 & 58.50 \\ \bottomrule\end{tabular}
\end{table}
\begin{figure}[htbp]\centering
\includegraphics[width=0.82\textwidth]{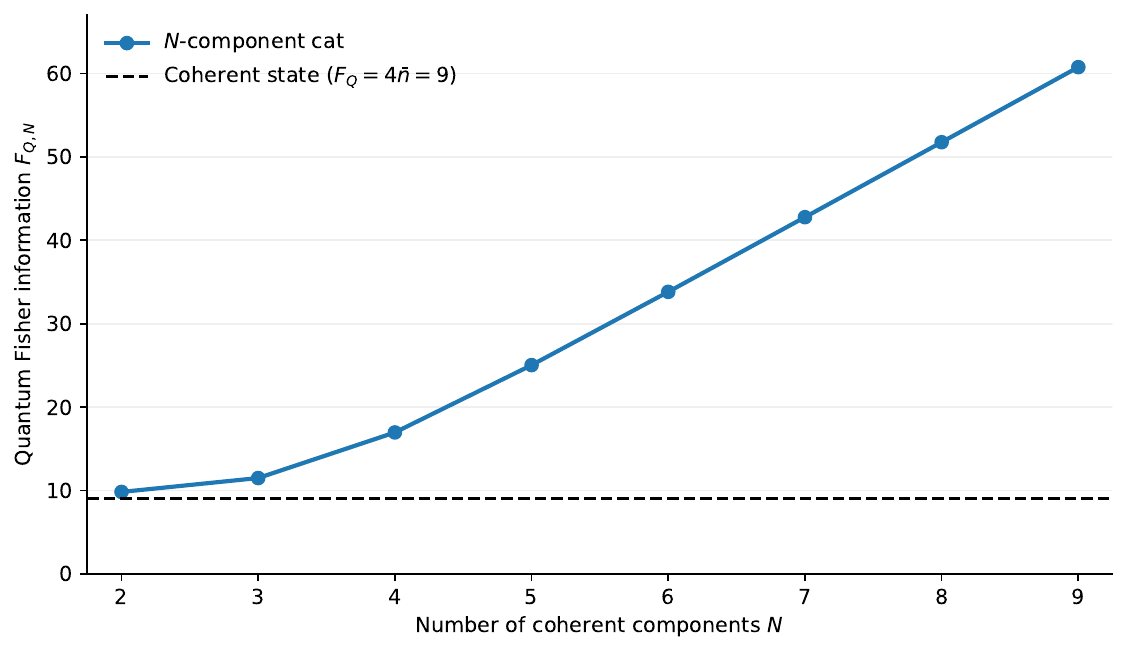}
\caption{Exact phase quantum Fisher information $F_{Q,N}$ for equal-phase cat states as a function of the component number $N=2,\ldots,9$ at the common mean photon number $\bar n=2.25$. The coherent amplitude $|\alpha_N|$ is independently adjusted for each $N$ using Eq.~\eqref{match}. The dashed line denotes the coherent-state benchmark $F_Q^{\mathrm{coh}}=4\bar n=9.00$. All values are evaluated from the exact expression~\eqref{fqexact}, valid for arbitrary $N$; the large-$N$ asymptote is given in Eq.~\eqref{qfi_linear}.}\label{fig:qfi}
\end{figure}

At fixed component amplitude $|\alpha|$, the selection of photon-number sectors $0,N,2N,\ldots$ changes the mean energy as $N$ varies. Solving Eq.~\eqref{match} instead fixes $\bar n$, so differences in QFI reflect photon-number variance at the same mean energy. This comparison does not match preparation costs, higher photon-number moments, phase-reference resources, or robustness to loss.

For an additional quantum benchmark, the optimal pure single-mode Gaussian probe for the generator $n$ at fixed mean photon number is squeezed vacuum~\cite{Monras2006}. Its QFI is
\begin{equation}\label{gaussianbenchmark}
F_Q^{\mathrm{sq\,vac}}=8\bar n(\bar n+1),
\qquad F_Q^{\mathrm{sq\,vac}}=58.50\quad(\bar n=2.25),
\end{equation}
with the fixed-energy optimization and derivation given in Ref.~\cite{Monras2006}, Eqs.~(22)--(25); see also Ref.~\cite{Lee2019}. Thus all cats in Table~\ref{tab:fixedmean} exceed the coherent-state benchmark, while only $N=9$ among those tabulated exceeds squeezed vacuum. This compares ideal local QFI at a common mean energy, without matching preparation or measurement resources. The phase-sensing comparisons in Refs.~\cite{Lee2015,Lee2020,Bibi2026} concern related probes and different generator or resource conventions; their numerical QFI values cannot be transferred directly to this table.

\paragraph{Large-$N$ behavior at fixed mean energy.}
The selection rule in Eq.~\eqref{fock} gives an exact decomposition
\begin{equation}\label{qfi_asymptotic}
F_{Q,N}=4\bar n(N-\bar n)+
4N^2\sum_{m=2}^{\infty}m(m-1)P_{Nm},
\end{equation}
where $P_{Nm}$ is the probability of $|Nm\rangle$. The nonnegative
remainder arises solely from the $2N,3N,\ldots$ sectors.
As shown in Appendix~\ref{app:largeN}, it vanishes at fixed
$\bar n>0$ as $N\to\infty$. Therefore
\begin{equation}\label{qfi_linear}
F_{Q,N}=4\bar n(N-\bar n)+o(1)
=9N-20.25+o(1)\quad(\bar n=2.25).
\end{equation}
This analytical asymptote, rather than a fit to Fig.~\ref{fig:qfi},
explains the near-linear trend. In particular,
$F_{Q,N}\ge4\bar n(N-\bar n)$ for every $N$.
The leading term in Eq.~\eqref{qfi_linear} equals the known QFI of $\sqrt{1-\bar n/N}\,|0\rangle+e^{i\vartheta_0}\sqrt{\bar n/N}\,|N\rangle$, for $N>\bar n$ and a fixed relative phase $\vartheta_0$~\cite{Lee2019}. Earlier circular-state work studied regimes approximating two-Fock-state superpositions~\cite{SouzaSilva2001,MizrahiJose2002}. Here the result concerns the full circular-cat photon-number distribution at fixed $\bar n$: Appendix~\ref{app:largeN} proves that the higher-sector contribution vanishes. It is therefore a realization of the established large-variance mechanism~\cite{Zhang2013,Lee2019} within this family, rather than a new general route around operational phase-estimation bounds.

For $N>\bar n$, the ideal quantum Cram\'er--Rao value obeys
\begin{equation}\label{qcrbbound}
\Delta\theta_{\rm QCRB}=\frac{1}{\sqrt{\nu F_{Q,N}}}
\leq\frac{1}{2\sqrt{\nu\bar n(N-\bar n)}},
\end{equation}
where $\nu$ denotes independent repetitions. This bounds the
\emph{ideal QCRB value}, not the uncertainty attained by an arbitrary
measurement. The large-$N$ limit assumes a lossless, phase-referenced
model and does not establish monotonicity for every finite $N$.
At fixed component amplitude $r$, by contrast, the state approaches
vacuum and both $\bar n_N(r)$ and $F_{Q,N}(r)$ vanish.

\section{Squeezing and local phase estimation}
Squeezed-cat probes have previously been investigated in two-mode phase estimation with explicit readout and loss models~\cite{Knott2016}. For the single-mode squeezed state $S(t)|\Psi_N\rangle$ and the generator $n$ used here, the exact phase QFI is
\begin{equation}\label{sqfq}
F_{Q,N}^{(S)}(t)=4\operatorname{Var}_{\Psi_N}[S^\dagger(t)nS(t)].
\end{equation}
The transformed number operator and the explicit formula in terms of photon-number and anomalous moments are derived in Appendix~\ref{app:sqQFI}. This QFI concerns a phase shift applied to the prepared squeezed state. It must be distinguished from estimating $\theta$ in the ordered $U(\theta)D(\delta)$ protocol when $\delta\ne0$; the generator variance is then evaluated in the displaced squeezed state. Since squeezing changes mean energy, claims of squeezing-induced metrological advantage require an additional resource-matched comparison.

\section{Quantum Cram\'er--Rao bound and physical interpretation}

For a measurement with outcome distribution $p(x|\theta)$,
the classical Fisher information is
\begin{equation}\label{cfi}
F_{\rm cl}(\theta)=
\sum_x p(x|\theta)
[\partial_\theta\ln p(x|\theta)]^2,
\end{equation}
where the sum is replaced by an integral for continuous outcomes.
For $\nu$ independent measurements and a locally unbiased estimator,
the classical and quantum Cram\'er--Rao bounds~\cite{BC,Paris} give
\begin{equation}\label{qcrb}
\operatorname{Var}(\widetilde\theta)
\geq \frac{1}{\nu F_{\rm cl}(\theta)}
\geq \frac{1}{\nu F_Q(\theta)},
\qquad
\Delta\theta\geq\frac{1}{\sqrt{\nu F_Q(\theta)}}.
\end{equation}
Thus, a larger QFI lowers the ideal bound on local phase-estimation uncertainty. By Eq.~\eqref{qcrbbound}, this bound tends to zero for equal-phase cats as $N\to\infty$ at fixed positive mean photon number.

This ideal limit does not guarantee experimentally attainable precision. Saturating the bound requires an appropriate measurement and a locally calibrated estimator~\cite{Paris}. For probes dominated by vacuum with rare high-photon-number components, rapidly varying phase dependence can coexist with low visibility and substantial requirements on prior phase information and sample size~\cite{Berry2012}. The relation between local sensitivity, estimator bias, and operational uncertainty is central to the discussion of sub-Heisenberg phase estimation~\cite{RivasLuis2012,Berry2012,Pezze2013}. Accordingly, sending $N$ to infinity in the QCRB alone does not establish a precision achievable with a fixed measurement budget or over a fixed prior interval.

Phase coherence between photon-number sectors requires an accessible phase reference~\cite{Jarzyna2012}. Without one, phase averaging gives a single-mode state diagonal in the photon-number basis; it commutes with $U(\theta)$ and therefore carries no information about this phase encoding. Equal-phase cat states also have a $2\pi/N$ rotational symmetry, evident from Eq.~\eqref{fock}, so the encoded state is periodic in $\theta$ and local QFI cannot resolve the resulting global ambiguity. Generic relative phases need not preserve this symmetry. Generator and reference conventions must be specified when comparing with two-mode schemes~\cite{Lee2015,Lee2020,Jarzyna2012}.

Finally, the present calculations assume pure states and unitary encoding. Photon loss imposes further limits that are not captured by the initial generator variance~\cite{Escher2011}; it must be analyzed for the actual prepared state and measurement. A reference beam, probabilistic preparation, and nonideal detection also contribute resources or errors beyond the mean probe energy~\cite{Alodjants2024,Jarzyna2012,Zeytinoglu2026}. None of the ideal QFI comparisons above constitutes a loss-tolerance or preparation-efficiency claim.

\section{Conclusion}
We have extended our earlier squeezed-cat overlap analysis to arbitrary component number and relative phases in a phase-referenced single-mode model. Building on the established circular-state construction and its sensing applications~\cite{Lee2015,Lee2020,Toscano2006,Shukla2023,Hailin2025,Bibi2026}, we evaluated the complex overlap under the ordered transformation $S^\dagger U(\theta)D(\delta)S$ with exact finite-amplitude normalization.

For equal-phase states at $\bar n=2.25$, the calculated QFI increases from $9.84$ for $N=2$ to $60.75$ for $N=9$. The same-energy coherent and squeezed-vacuum values are $9.00$ and $58.50$, respectively; only the $N=9$ state among the tabulated cats exceeds both benchmarks. We also established the fixed-energy asymptote $F_{Q,N}=4\bar n(N-\bar n)+o(1)$ by bounding the higher photon-number sectors. Its leading term agrees with the known vacuum--$N$-photon result~\cite{Lee2019}; the tail estimate establishes its applicability to the circular-cat family. At fixed positive mean photon number, the corresponding ideal local QCRB tends to zero as $N\to\infty$, subject to the reference and statistical qualifications discussed above~\cite{Berry2012,Jarzyna2012}.

Applying these ideal-state results to a physical sensor requires the QFI of the actually prepared state, together with its preparation probability, phase reference, losses, and readout.

\section*{Acknowledgments}
R. Singh thanks Prof. F. Ya. Khalili for suggesting the investigation
of displacement-induced orthogonality in multicomponent Schr\"odinger cat states.
\appendix
\section{Normalization and photon-number moments}\label{app:norm}
For completeness, we reproduce the circular-state selection rule and normalization used in Refs.~\cite{MizrahiJose2002,LanXu2022}. Insert $|\beta\rangle=e^{-|\beta|^2/2}\sum_{n\ge0}
\beta^n|n\rangle/\sqrt{n!}$ into Eq.~\eqref{state}.
For $\phi_k=0$, the root-of-unity sum
$\sum_{k=0}^{N-1}\zeta_N^{kn}=N$ if $N$ divides $n$ and zero
otherwise yields the Fock expansion~\eqref{fock}. Its unnormalized
squared norm is $N^2e^{-r}Z_N(r)$; consequently
$p_{Nm}=r^{Nm}/[(Nm)!Z_N(r)]$ and $p_n=0$ otherwise.
Termwise differentiation gives
\begin{equation}\label{moments}
\langle n\rangle=r\frac{Z_N'}{Z_N},\quad
\langle n(n-1)\rangle=r^2\frac{Z_N''}{Z_N},\quad
\operatorname{Var}(n)=r^2\frac{Z_N''}{Z_N}
+r\frac{Z_N'}{Z_N}-r^2\left(\frac{Z_N'}{Z_N}\right)^2.
\end{equation}
The root-of-unity form gives
$Z_N^{(q)}(r)=N^{-1}\sum_k\zeta_N^{kq}e^{r\zeta_N^k}$.
For fixed $N$ and $r>0$, differentiation of the mean gives
\begin{equation*}
\frac{d\bar n_N(r)}{dr}=\frac{\operatorname{Var}_{\Psi_N}(n)}{r}>0.
\end{equation*}
The variance is strictly positive because both $P_0$ and $P_N$ are nonzero. The Fock series gives $\bar n_N(r)\to0$ as $r\to0$. In the root-of-unity sum, the term $e^r$ dominates as $r\to\infty$, so $Z_N(r)\sim e^r/N$ and $\bar n_N(r)\sim r\to\infty$. Thus every prescribed $\bar n>0$ has exactly one positive solution $r_N$. Solving Eq.~\eqref{match} gives Table~\ref{tab:fixedmean}.

For arbitrary $\phi_k$, the coefficient of $|n\rangle$ is
$\mathcal N_N e^{-r/2}\alpha^n
\sum_k e^{i\phi_k}\zeta_N^{kn}/\sqrt{n!}$;
its normalized squared modulus $P_n$ independently yields
$F_{Q,N}=4[\sum_n n^2P_n-(\sum_n nP_n)^2]$.
Unlike the equal-phase case, generic phases need not select $n=0\bmod N$.
\section{Proof of the fixed-energy asymptote}\label{app:largeN}
With $P_{Nm}=r_N^{Nm}/[(Nm)!Z_N(r_N)]$ and
$\sum_m NmP_{Nm}=\bar n$, the identity
$(Nm)^2=N(Nm)+N^2m(m-1)$ gives
Eq.~\eqref{qfi_asymptotic}. To control its remainder, set
$t_N=r_N^N/N!=P_N/P_0$. For $N>\bar n$,
\begin{equation*}
P_0\ge1-\bar n/N,\qquad P_N\le\bar n/N,
\qquad t_N\le\frac{\bar n}{N-\bar n}.
\end{equation*}
Successive probabilities satisfy
\begin{equation}\label{tailratio}
\frac{P_{(m+1)N}}{P_{mN}}
=t_N\frac{N!(mN)!}{((m+1)N)!}
\le\rho_N:=
\frac{\bar n}{(N-\bar n)\binom{2N}{N}}
\quad(m\ge1).
\end{equation}
Thus $P_{Nm}\le P_N\rho_N^{m-1}$ and, for $\rho_N<1$,
\begin{equation*}
0\le4N^2\sum_{m\ge2}m(m-1)P_{Nm}
\le\frac{8N\bar n\,\rho_N}{(1-\rho_N)^3}
\longrightarrow0.
\end{equation*}
The last inequality uses
$\sum_{m\ge2}m(m-1)\rho^{m-1}
=2\rho/(1-\rho)^3$ and $NP_N\le\bar n$.
Since $\binom{2N}{N}$ grows exponentially, the correction
vanishes, proving Eq.~\eqref{qfi_linear}.

\section{Coherent-state derivation of the finite-transformation kernel}\label{app:gaussian}
Displacement, squeezing, and phase rotation are standard Gaussian unitaries~\cite{Weedbrook2012}. We evaluate their ordered coherent-state matrix element directly, retaining its complex phase, without reconstructing the composite unitary solely from a Bogoliubov transformation. Throughout, $a|z\rangle=z|z\rangle$, $\langle z|w\rangle=\exp[-(|z|^2+|w|^2)/2+z^*w]$, and
\begin{equation}\label{cohidentity}
I=\int_{\mathbb C}\frac{d^2z}{\pi}|z\rangle\langle z|.
\end{equation}
We retain the ordering $S^\dagger U D S$ and the squeezing Hamiltonian
\begin{equation}\label{squeeze}
S(t)=\exp[-igt(A_2^*a^2+A_2a^{\dagger2})],\qquad
U(\theta)=e^{-i\theta a^\dagger a},\qquad D(\delta)=e^{\delta a^\dagger-\delta^*a}.
\end{equation}
Write $A_2=|A_2|e^{i\varphi}$ and $\tau=2g|A_2|t$, with $\mu=\cosh\tau$. Define the coefficients
\begin{equation}\label{integralparams}
q=-ie^{i\varphi}\tanh\tau,\quad b=\operatorname{sech}\tau,\quad h=e^{-i\theta},\quad \Delta=1-|q|^2h^2.
\end{equation}
For $A_2=0$ take $q=0$, $b=1$. Since $|q|<1$ for finite real $\tau$, $\Delta\ne0$. The square root $\sqrt\Delta$ is chosen by continuous continuation from $\tau=0$, where it equals $1$.

To derive the squeezed coherent-state kernel, use the factorization
\begin{equation}\label{squeezefactorization}
S(t)=\exp\!\left(\frac q2a^{\dagger2}\right)
\exp\!\left[-(\ln\mu)\left(n+\frac12\right)\right]
\exp\!\left(-\frac{q^*}{2}a^2\right).
\end{equation}
Indeed, $[n,a^{\dagger2}]=2a^{\dagger2}$ and $[a^2,a^{\dagger2}]=4n+2$ show that the right-hand side of Eq.~\eqref{squeezefactorization} satisfies $i\partial_t S=g(A_2^*a^2+A_2a^{\dagger2})S$ and equals $I$ at $t=0$, which fixes the factorization. In its coherent-state matrix element, the outer factors give $\exp[q(z^*)^2/2-q^*w^2/2]$, while the middle factor gives $\mu^{-1/2}\exp[-(|z|^2+|w|^2)/2+z^*w/\mu]$. Since $b=1/\mu$, their product is
\begin{equation}\label{Skernel}
\langle z|S(t)|w\rangle=\frac{e^{-(|z|^2+|w|^2)/2}}{\sqrt\mu}
\exp\!\left[\frac q2(z^*)^2+bz^*w-\frac{q^*}{2}w^2\right].
\end{equation}
The other two kernels, with their phases fixed, are
\begin{align}\label{UDkernels}
\langle z|U(\theta)|w\rangle&=e^{-(|z|^2+|w|^2)/2+hz^*w},\\
\langle z|D(\delta)|w\rangle&=\exp\!\left[-\frac{|z|^2+|w|^2+|\delta|^2}{2}+z^*(w+\delta)-\delta^*w\right].
\end{align}
For arbitrary $x,y\in\mathbb C$, insert three coherent-state resolutions of identity:
\begin{align}\label{tripleintegral}
\chi(x,y;\theta,\delta,t)&=\langle x|S^\dagger UDS|y\rangle\\
&=\int\frac{d^2z_1\,d^2z_2\,d^2z_3}{\pi^3}
\langle x|S^\dagger|z_1\rangle\langle z_1|U|z_2\rangle
\langle z_2|D|z_3\rangle\langle z_3|S|y\rangle.\nonumber
\end{align}
Using the completeness relation once to combine $U$ and $D$ gives the exact identity
\begin{equation}\label{UDcombined}
\langle z|UD|w\rangle=\exp\!\left[-\frac{|z|^2+|w|^2+|\delta|^2}{2}
+h z^*(w+\delta)-\delta^*w\right].
\end{equation}
Consequently the triple integral reduces to two complex Gaussian integrations:
\begingroup\small
\begin{align}\label{doubleintegral}
\chi(x,y;\theta,\delta,t)
={}&\frac{e^{-\frac12(|x|^2+|y|^2+|\delta|^2)
-\frac q2(x^*)^2-\frac{q^*}{2}y^2}}{\mu}\\
&\times\int_{\mathbb C^2}\frac{d^2z\,d^2w}{\pi^2}
\exp\Big[-|z|^2-|w|^2+\frac{q^*}{2}z^2+b x^*z\\
&\hspace{25mm}+h z^*(w+\delta)+\frac q2(w^*)^2+b w^*y-\delta^*w\Big].\nonumber
\end{align}
\endgroup
The elementary complex Gaussian identities used here are
\begin{align}\label{gaussidentities}
\int_{\mathbb C}{d^2z}/{\pi}e^{-|z|^2+\frac A2z^2+Bz+Cz^*}
&=e^{\frac A2C^2+BC},\\
\int_{\mathbb C}{d^2w}/{\pi}e^{-|w|^2+\frac A2w^2+\frac B2(w^*)^2+Cw+Dw^*}
&={1}/{\sqrt{1-AB}}
\exp\left[\frac{AD^2+BC^2+2CD}{2(1-AB)}\right],\nonumber
\end{align}
with the branch obtained by continuous deformation from $A=B=0$; the parameters encountered in our integrals satisfy the convergence requirements. Integrating over $z$ first, and then $w$, define
\begin{equation}\label{kernelC}
C(x)=q^*h^2\delta+bhx^*-\delta^*.
\end{equation}
The exact result is
\begin{equation}\label{generalK}
\begin{aligned}
\chi(x,y;\theta,\delta,t)={1}/{(\mu\sqrt\Delta)}\exp\Bigg\{&-\frac{|x|^2+|y|^2+|\delta|^2}{2}
-\frac q2(x^*)^2-\frac{q^*}{2}y^2
+\frac{q^*h^2}{2}\delta^2+bhx^*\delta\\
&+\frac{q^*h^2b^2y^2+q[C(x)]^2+2byC(x)}{2\Delta}\Bigg\}.
\end{aligned}
\end{equation}
Unlike a kernel reconstructed only from the transformed annihilation operator, Eq.~\eqref{generalK} contains the full complex phase fixed by the ordered operator product. For arbitrary phases the normalized cat squared state overlap is Eq.~\eqref{generalphasegamma}. For equal phases, it reduces to
\begin{equation}\label{fullgamma}
\gamma_N(\theta,\delta,t)=\mathcal N_N^4
\left|\sum_{j,k=0}^{N-1}\chi(\alpha_j,\alpha_k;\theta,\delta,t)\right|^2,
\qquad \alpha_k=\alpha e^{2\pi i k/N}.
\end{equation}
Eq.~\eqref{generalphasegamma} involves $N^2$ terms and no numerical integration or Fock-space cutoff.

\section{Limiting cases and displacement orthogonality}\label{app:cases}
Displacement overlaps for circular and generalized compass states were analyzed in Refs.~\cite{Toscano2006,Shukla2023,Hailin2025}. The general kernel~\eqref{generalK} gives the following limits, with finite-amplitude normalization retained.
For $t=0$, one has $q=0$, $b=\Delta=1$, hence
\begin{equation}\label{nosqueeze}
\chi(x,y;\theta,\delta,0)=
\exp\!\left[\frac{\delta y^*-\delta^*y}{2}
-\frac{|x|^2+|y+\delta|^2}{2}
+x^*(y+\delta)e^{-i\theta}\right].
\end{equation}
Setting also $\delta=0$ gives the pure-rotation overlap for
equal-phase states:
\begin{equation}\label{phaseonly}
\gamma_N(\theta,0,0)=
\left|\frac{Z_N(re^{-i\theta})}{Z_N(r)}\right|^2,
\quad Z_N(z)=\sum_{m\ge0}\frac{z^{Nm}}{(Nm)!}.
\end{equation}
For pure displacement ($\theta=t=0$), define
\begin{equation}\label{Cdef}
\mathcal C_N(\eta)=\mathcal N_N^2
\sum_{j,k=0}^{N-1}e^{i(\phi_k-\phi_j)}
\exp\!\left[\frac{\eta\alpha_k^*-\eta^*\alpha_k}{2}
-\frac{|\alpha_j|^2+|\alpha_k+\eta|^2}{2}
+\alpha_j^*(\alpha_k+\eta)\right].
\end{equation}
The squeeze-conjugation identity follows from
$S^\dagger aS=\mu a+\nu a^\dagger$,
$\mu=\cosh\tau$, $\nu=-ie^{i\varphi}\sinh\tau$:
\begin{equation}\label{sqd}
S^\dagger D(\delta)S=D(\mu\delta-\nu\delta^*),
\qquad
\gamma_N(0,\delta,t)
=|\mathcal C_N(\mu\delta-\nu\delta^*)|^2.
\end{equation}
Thus ordinary and squeezed cats are orthogonal to their displaced
counterparts precisely when, respectively,
\begin{equation}\label{SCSSCzeros}
\mathcal C_N(\delta)=0,\qquad
\mathcal C_N(\mu\delta-\nu\delta^*)=0.
\end{equation}
The map $\delta\mapsto\mu\delta-\nu\delta^*$ is invertible because
$|\mu|^2-|\nu|^2=1$; squeezing changes the directional displacement
scale but need not improve it in every direction. When exact zeros
exist, the smallest nonzero $|\delta|$ satisfying the relevant
condition defines the orthogonality displacement. For approximate
orthogonality, a common nonzero threshold on $\gamma_N$ must be
specified. For $\delta=0$, Eq.~\eqref{generalK} describes squeezed
phase rotation; at $\theta=\delta=0$ it reduces to unit overlap.
For vacuum input and no squeezing,
$\gamma_{\rm vac}(0,\delta,0)=e^{-|\delta|^2}$, illustrating that
an exact zero need not exist at finite displacement.
\section{Derivation of squeezed-state phase QFI}\label{app:sqQFI}
With $S(t)$ as defined in Appendix~\ref{app:gaussian}, for $G_n=S^\dagger nS$, write $c=|\mu|^2+|\nu|^2$, $b=|\nu|^2$, $h=\mu^*\nu$. Then
\begin{equation}\label{qfiExpanded}
F_{Q,N}^{(S)}=4\left[\langle(c n+b+h a^{\dagger2}+h^*a^2)^2\rangle_{\Psi_N}
-\langle c n+b+h a^{\dagger2}+h^*a^2\rangle_{\Psi_N}^{\,2}\right].
\end{equation}
Eq.~\eqref{qfiExpanded} holds for arbitrary component phases, with moments obtained from the coefficients in Appendix~\ref{app:norm}. In the remainder of this appendix we restrict to equal phases, $\phi_k=0$, and use the Fock expansion~\eqref{fock}. For nonnegative integers $p,q$, this gives
\begin{equation}\label{amom}
\langle(a^\dagger)^p a^q\rangle_{\Psi_N}=
\frac{1}{Z_N(r)}\sum_{m,l\ge0}\frac{(\alpha^*)^{Nl}\alpha^{Nm}}{\sqrt{(Nl)!(Nm)!}}
\langle Nl|(a^\dagger)^p a^q|Nm\rangle.
\end{equation}
The nonzero terms satisfy $p-q\equiv0\pmod N$. In particular $\langle a^2\rangle=0$ for $N\ge3$, while it may be nonzero for $N=2$; this distinction affects the squeezing-angle dependence. These formulas do not assume that squeezing improves precision at fixed mean photon number.

\paragraph{Closed squeezed-state QFI for rotational cats.}
For the equal-phase state, define $m_1=\langle n\rangle$, $m_2=\langle n^2\rangle$, $V=m_2-m_1^2$, and
\begin{equation}\label{qws}
q_N=\langle a^2\rangle=\begin{cases}\alpha^2,&N=2,\\0,&N\ge3,\end{cases}
\qquad
w_N=\langle a^4\rangle=\begin{cases}\alpha^4,&N=2\text{ or }4,\\0,&\text{otherwise}.\end{cases}
\end{equation}
The state obeys $a^N|\Psi_N\rangle=\alpha^N|\Psi_N\rangle$. Using $c=|\mu|^2+|\nu|^2$ and $h=\mu^*\nu$, the exact squeezed phase QFI takes the explicit form
\begin{equation}\label{sqfqclosed}
\begin{aligned}
\frac{F_{Q,N}^{(S)}}4={}&c^2 V+4c\operatorname{Re}(h^*q_N)\\
&+2|h|^2(m_2+m_1+1-|q_N|^2)\\
&+2\operatorname{Re}\!\left[(h^*)^2(w_N-q_N^2)\right].
\end{aligned}
\end{equation}
Here $m_1=rZ_N'/Z_N$ and $m_2=r^2Z_N''/Z_N+m_1$ are exact. At $\alpha=1.5$ one sets $r=2.25$, $q_2=2.25$, $w_2=w_4=5.0625$, and all other $q_N,w_N$ vanish for $N=3,4$ as prescribed by Eq.~\eqref{qws}. Eq.~\eqref{sqfqclosed} applies for arbitrary real $t$ without a small-squeezing expansion and reduces to $4V$ at $t=0$.


\begin{thebibliography}{99}
\bibitem{Dodonov1974}V.~V.~Dodonov, I.~A.~Malkin, and V.~I.~Man'ko, ``Even and odd coherent states and excitations of a singular oscillator,'' \emph{Physica} \textbf{72}, 597--615 (1974). doi:\href{https://doi.org/10.1016/0031-8914(74)90215-8}{\nolinkurl{10.1016/0031-8914(74)90215-8}}.

\bibitem{Janszky1993}J.~Janszky, P.~Domokos, and P.~Adam, ``Coherent states on a circle and quantum interference,'' \emph{Phys. Rev. A} \textbf{48}, 2213--2219 (1993). doi:\href{https://doi.org/10.1103/physreva.48.2213}{\nolinkurl{10.1103/physreva.48.2213}}.

\bibitem{MizrahiJose2002}S.~S.~Mizrahi and W.~D.~Jos\'e, ``Generating Fock states and two-Fock states superposition from circular states, in a trapped ion,'' contribution to the VII International Conference on Squeezed States and Uncertainty Relations (Boston, 2001), \href{https://arxiv.org/abs/quant-ph/0210177}{arXiv:quant-ph/0210177} (2002).

\bibitem{LanXu2022}B.~Lan and X.-x.~Xu, ``Multi-Headed Symmetrical Superpositions of Coherent States,'' \emph{Int. J. Theor. Phys.} \textbf{61}, 148 (2022). doi:\href{https://doi.org/10.1007/s10773-022-05134-6}{\nolinkurl{10.1007/s10773-022-05134-6}}. \href{https://arxiv.org/abs/2109.08056}{arXiv:2109.08056}.

\bibitem{ST}R.~Singh and A.~E.~Teretenkov, ``Quantum sensitivity of squeezed Schrodinger cat states,'' \emph{Physics Open} \textbf{18}, 100198 (2024). doi:\href{https://doi.org/10.1016/j.physo.2023.100198}{\nolinkurl{10.1016/j.physo.2023.100198}}.

\bibitem{Alodjants2024}A.~P.~Alodjants, D.~V.~Tsarev, D.~A.~Kuts, S.~A.~Podoshvedov, and S.~P.~Kulik, ``Quantum optical metrology,'' \emph{Physics-Uspekhi} \textbf{67}, 668--693 (2024). doi:\href{https://doi.org/10.3367/ufne.2024.01.039634}{\nolinkurl{10.3367/ufne.2024.01.039634}}.

\bibitem{Zurek2001}W.~H.~Zurek, ``Sub-Planck structure in phase space and its relevance for quantum decoherence,'' \emph{Nature} \textbf{412}, 712--717 (2001). doi:\href{https://doi.org/10.1038/35089017}{\nolinkurl{10.1038/35089017}}. \href{https://arxiv.org/abs/quant-ph/0201118}{arXiv:quant-ph/0201118}.

\bibitem{Toscano2006}F.~Toscano, D.~A.~R.~Dalvit, L.~Davidovich, and W.~H.~Zurek, ``Sub-Planck phase-space structures and Heisenberg-limited measurements,'' \emph{Phys. Rev. A} \textbf{73}, 023803 (2006). doi:\href{https://doi.org/10.1103/physreva.73.023803}{\nolinkurl{10.1103/physreva.73.023803}}.

\bibitem{Shukla2023}A.~Shukla and B.~C.~Sanders, ``Superposing compass states for asymptotic isotropic sub-Planck phase-space sensitivity,'' \emph{Phys. Rev. A} \textbf{108}, 043719 (2023). doi:\href{https://doi.org/10.1103/physreva.108.043719}{\nolinkurl{10.1103/physreva.108.043719}}. \href{https://arxiv.org/abs/2306.13182}{arXiv:2306.13182}.

\bibitem{Hailin2025}Tan Hailin, Naeem Akhtar, and Gao Xianlong, ``Multicomponent cat states with sub-Planckian structures and their optomechanical analogues,'' \emph{Phys. Rev. Applied} \textbf{24}, 024053 (2025). doi:\href{https://doi.org/10.1103/gpj3-bbxq}{\nolinkurl{10.1103/gpj3-bbxq}}. \href{https://arxiv.org/abs/2411.13349}{arXiv:2411.13349}.

\bibitem{Lee2015}S.-Y.~Lee, C.-W.~Lee, H.~Nha, and D.~Kaszlikowski, ``Quantum phase estimation using a multi-headed cat state,'' \emph{J. Opt. Soc. Am. B} \textbf{32}, 1186--1192 (2015). doi:\href{https://doi.org/10.1364/josab.32.001186}{\nolinkurl{10.1364/josab.32.001186}}. \href{https://arxiv.org/abs/1409.4220}{arXiv:1409.4220}.

\bibitem{Lee2020}S.-W.~Lee, S.-Y.~Lee, and J.~Kim, ``Optimal quantum phase estimation with generalized multi-component Schr\"odinger cat states,'' \emph{J. Opt. Soc. Am. B} \textbf{37}, 2423--2429 (2020). doi:\href{https://doi.org/10.1364/josab.393200}{\nolinkurl{10.1364/josab.393200}}. \href{https://arxiv.org/abs/2003.06302}{arXiv:2003.06302}.

\bibitem{Bibi2026}T.~Bibi, S.~Javed, A.~Naeem, and S.~Iqbal, ``Multiheaded Schr\"odinger Cats and their Power for Quantum Metrology,'' \emph{J. Russ. Laser Res.} \textbf{47}, 146--160 (2026). doi:\href{https://doi.org/10.1007/s10946-026-10287-4}{\nolinkurl{10.1007/s10946-026-10287-4}}.

\bibitem{BibiIqbal2026}T.~Bibi and S.~Iqbal, ``Quantum-enhanced phase estimation with photon-engineered multi-headed cat states,'' \emph{Opt. Express} \textbf{34}, 27440--27459 (2026). doi:\href{https://doi.org/10.1364/oe.605086}{\nolinkurl{10.1364/oe.605086}}.

\bibitem{Zhang2013}Y.~R.~Zhang, G.~R.~Jin, J.~P.~Cao, W.~M.~Liu, and H.~Fan, ``Unbounded quantum Fisher information in two-path interferometry with finite photon number,'' \emph{J. Phys. A: Math. Theor.} \textbf{46}, 035302 (2013). doi:\href{https://doi.org/10.1088/1751-8113/46/3/035302}{\nolinkurl{10.1088/1751-8113/46/3/035302}}. \href{https://arxiv.org/abs/1105.2990}{arXiv:1105.2990}.

\bibitem{Lee2019}C.~Lee, C.~Oh, H.~Jeong, C.~Rockstuhl, and S.-Y.~Lee, ``Using states with a large photon number variance to increase quantum Fisher information in single-mode phase estimation,'' \emph{J. Phys. Commun.} \textbf{3}, 115008 (2019). doi:\href{https://doi.org/10.1088/2399-6528/ab524a}{\nolinkurl{10.1088/2399-6528/ab524a}}. \href{https://arxiv.org/abs/1907.06349}{arXiv:1907.06349}.

\bibitem{RivasLuis2012}\'A.~Rivas and A.~Luis, ``Sub-Heisenberg estimation of non-random phase shifts,'' \emph{New J. Phys.} \textbf{14}, 093052 (2012). doi:\href{https://doi.org/10.1088/1367-2630/14/9/093052}{\nolinkurl{10.1088/1367-2630/14/9/093052}}. \href{https://arxiv.org/abs/1105.6310}{arXiv:1105.6310}.

\bibitem{Berry2012}D.~W.~Berry, M.~J.~W.~Hall, M.~Zwierz, and H.~M.~Wiseman, ``Optimal Heisenberg-style bounds for the average performance of arbitrary phase estimates,'' \emph{Phys. Rev. A} \textbf{86}, 053813 (2012). doi:\href{https://doi.org/10.1103/physreva.86.053813}{\nolinkurl{10.1103/physreva.86.053813}}. \href{https://arxiv.org/abs/1209.3547}{arXiv:1209.3547}.

\bibitem{Pezze2013}L.~Pezz\'e, ``Sub-Heisenberg phase uncertainties,'' \emph{Phys. Rev. A} \textbf{88}, 060101(R) (2013). doi:\href{https://doi.org/10.1103/physreva.88.060101}{\nolinkurl{10.1103/physreva.88.060101}}.

\bibitem{STPLA}R.~Singh and A.~E.~Teretenkov, ``The formation of entangled Schr\"odinger cat-like states in the process of spontaneous parametric down-conversion,'' \emph{Phys. Lett. A} \textbf{574}, 131382 (2026). doi:\href{https://doi.org/10.1016/j.physleta.2026.131382}{\nolinkurl{10.1016/j.physleta.2026.131382}}. \href{https://arxiv.org/abs/2405.14526}{arXiv:2405.14526}.

\bibitem{SinghTeretenkov2026Microring}R.~Singh and A.~E.~Teretenkov, ``Generation of Schr\"odinger cat-like states via degenerate dual pump spontaneous four-wave mixing in a $\chi^{(3)}$ microring resonator,'' \href{https://arxiv.org/abs/2604.15275}{arXiv:2604.15275} (2026). doi:\href{https://doi.org/10.48550/arXiv.2604.15275}{\nolinkurl{10.48550/arXiv.2604.15275}}.

\bibitem{SinghTeretenkov2026Kerr}R.~Singh and A.~E.~Teretenkov, ``Quantum pump depletion and multicomponent Schr\"odinger-cat-like states in doubly pumped intraresonance Kerr microresonators,'' \href{https://arxiv.org/abs/2606.19085}{arXiv:2606.19085} (2026). doi:\href{https://doi.org/10.48550/arXiv.2606.19085}{\nolinkurl{10.48550/arXiv.2606.19085}}.

\bibitem{Vlastakis2013}B.~Vlastakis, G.~Kirchmair, Z.~Leghtas, S.~E.~Nigg, L.~Frunzio, S.~M.~Girvin, M.~Mirrahimi, M.~H.~Devoret, and R.~J.~Schoelkopf, ``Deterministically Encoding Quantum Information Using 100-Photon Schr\"odinger Cat States,'' \emph{Science} \textbf{342}, 607--610 (2013). doi:\href{https://doi.org/10.1126/science.1243289}{\nolinkurl{10.1126/science.1243289}}.

\bibitem{Zeytinoglu2026}S.~Zeytino\u{g}lu, ``Generalized parity measurements and efficient large multicomponent cat state preparation with quantum signal processing,'' \emph{Phys. Rev. Research} \textbf{8}, 023208 (2026). doi:\href{https://doi.org/10.1103/xsgf-txnq}{\nolinkurl{10.1103/xsgf-txnq}}. \href{https://arxiv.org/abs/2409.05186}{arXiv:2409.05186}.

\bibitem{Monras2006}A.~Monras, ``Optimal phase measurements with pure Gaussian states,'' \emph{Phys. Rev. A} \textbf{73}, 033821 (2006). doi:\href{https://doi.org/10.1103/physreva.73.033821}{\nolinkurl{10.1103/physreva.73.033821}}. \href{https://arxiv.org/abs/quant-ph/0509018}{arXiv:quant-ph/0509018}.

\bibitem{BC}S.~L.~Braunstein and C.~M.~Caves, ``Statistical distance and the geometry of quantum states,'' \emph{Phys. Rev. Lett.} \textbf{72}, 3439--3443 (1994). doi:\href{https://doi.org/10.1103/physrevlett.72.3439}{\nolinkurl{10.1103/physrevlett.72.3439}}.

\bibitem{Paris}M.~G.~A.~Paris, ``Quantum estimation for quantum technology,'' \emph{Int. J. Quantum Inf.} \textbf{7}, Suppl.~1, 125--137 (2009). doi:\href{https://doi.org/10.1142/s0219749909004839}{\nolinkurl{10.1142/s0219749909004839}}. \href{https://arxiv.org/abs/0804.2981}{arXiv:0804.2981}.

\bibitem{SouzaSilva2001}A.~L.~Souza Silva, W.~Duarte Jos\'e, V.~V.~Dodonov, and S.~S.~Mizrahi, ``Production of two-Fock states superpositions from even circular states and their decoherence,'' \emph{Phys. Lett. A} \textbf{282}, 235--244 (2001). doi:\href{https://doi.org/10.1016/s0375-9601(01)00195-5}{\nolinkurl{10.1016/s0375-9601(01)00195-5}}.

\bibitem{Knott2016}P.~A.~Knott, T.~J.~Proctor, A.~J.~Hayes, J.~P.~Cooling, and J.~A.~Dunningham, ``Practical quantum metrology with large precision gains in the low-photon-number regime,'' \emph{Phys. Rev. A} \textbf{93}, 033859 (2016). doi:\href{https://doi.org/10.1103/physreva.93.033859}{\nolinkurl{10.1103/physreva.93.033859}}. \href{https://arxiv.org/abs/1505.04011}{arXiv:1505.04011}.

\bibitem{Jarzyna2012}M.~Jarzyna and R.~Demkowicz-Dobrza\'nski, ``Quantum interferometry with and without an external phase reference,'' \emph{Phys. Rev. A} \textbf{85}, 011801(R) (2012). doi:\href{https://doi.org/10.1103/physreva.85.011801}{\nolinkurl{10.1103/physreva.85.011801}}. \href{https://arxiv.org/abs/1108.3844}{arXiv:1108.3844}.

\bibitem{Escher2011}B.~M.~Escher, R.~L.~de Matos Filho, and L.~Davidovich, ``General framework for estimating the ultimate precision limit in noisy quantum-enhanced metrology,'' \emph{Nat. Phys.} \textbf{7}, 406--411 (2011). doi:\href{https://doi.org/10.1038/nphys1958}{\nolinkurl{10.1038/nphys1958}}. \href{https://arxiv.org/abs/1201.1693}{arXiv:1201.1693}.

\bibitem{Weedbrook2012}C.~Weedbrook, S.~Pirandola, R.~Garc\'ia-Patr\'on, N.~J.~Cerf, T.~C.~Ralph, J.~H.~Shapiro, and S.~Lloyd, ``Gaussian quantum information,'' \emph{Rev. Mod. Phys.} \textbf{84}, 621--669 (2012). doi:\href{https://doi.org/10.1103/revmodphys.84.621}{\nolinkurl{10.1103/revmodphys.84.621}}. \href{https://arxiv.org/abs/1110.3234}{arXiv:1110.3234}.
\end{thebibliography}
\end{document}